\documentclass[letterpaper, 10 pt, conference]{ieeeconf}  

\IEEEoverridecommandlockouts                              

\usepackage{graphicx}
\usepackage{picinpar}
\usepackage{amsmath}
\usepackage{amssymb}
\usepackage{url}
\usepackage{flushend}
\usepackage[latin1]{inputenc}
\usepackage{colortbl}
\usepackage{soul}
\usepackage{multirow}
\usepackage{pifont}
\usepackage{bm}
\usepackage{color}
\usepackage{alltt}
\usepackage[hidelinks]{hyperref}
\usepackage{enumerate}
\usepackage{siunitx}
\usepackage{breakurl}
\usepackage{epstopdf}
\usepackage{pbox}
\usepackage{booktabs}
\usepackage{subfig}
\usepackage{balance}
\usepackage{graphicx}
\usepackage{lipsum}
\usepackage{textcomp,mathcomp}
\usepackage[noadjust]{cite}
\newcommand{\norm}[1]{\left\lVert#1\right\rVert}

\usepackage{bm}

\usepackage{epstopdf}
\title{\LARGE \bf
Efficient Fault Diagnosis in Lithium-Ion Battery Packs: A Structural Approach with Moving Horizon Estimation}

\author{Amir Farakhor$^{1}$, Di Wu$^{2}$, Yebin Wang$^{3}$ and Huazhen Fang$^{1}$
\thanks{$^{1}$A. Farakhor and H. Fang are with the Department of Mechanical Engineering, Univeristy of Kansas, Lawrence, KS, USA.
        {\tt\small a.farakhor@ku.edu, fang@ku.edu}}%
\thanks{$^{2}$D. Wu is with the Pacific Northwest National Laboratory, Richland, WA, USA.
        {\tt\small di.wu@pnnl.gov}}
\thanks{$^{3}$Y. Wang is with the Mitsubishi Electric Research Laboratories, Cambridge, MA, USA.
        {\tt\small yebinwang@ieee.org}}%
}

\begin{document}

\maketitle
\thispagestyle{empty}
\pagestyle{empty}

\begin{abstract}
Safe and reliable operation of lithium-ion battery packs depends on effective fault diagnosis. However, model-based approaches often encounter two major challenges: high computational complexity and extensive sensor requirements. To address these bottlenecks, this paper introduces a novel approach that harnesses the structural properties of battery packs, including cell uniformity and the sparsity of fault occurrences. We integrate this approach into a Moving Horizon Estimation (MHE) framework and estimate fault signals such as internal and external short circuits and faults in voltage and current sensors. To mitigate computational demands, we propose a hierarchical solution to the MHE problem. The proposed solution breaks up the pack-level MHE problem into smaller problems and solves them efficiently. Finally, we perform extensive simulations across various battery pack configurations and fault types to demonstrate the effectiveness of the proposed approach. The results highlight that the proposed approach simultaneously reduces the computational demands and sensor requirements of fault diagnosis.
\end{abstract}
\section{INTRODUCTION}
Lithium-ion battery systems are pivotal to the electrification of transportation and the transition towards energy decarbonization. Their success can be attributed to key characteristics such as high energy density, long cycle life, and low self-discharge rates \cite{Wiley-MR-2024}. However, lithium-ion batteries also present significant safety risks due to the possibility of thermal runaway, which can lead to fires or even explosions \cite{RenewableEnergyReview-DH-2024}. Several well-publicized incidents have highlighted these safety concerns, particularly as lithium-ion batteries are increasingly used in safety-critical applications, such as electric aircraft \cite{Combustion-PP-2024, SNAME-KI-2024}. To mitigate these risks, battery management systems (BMS) are equipped with advanced fault diagnosis capabilities. A primary function of BMS is to monitor cell behavior via sensor measurements and take appropriate actions in the event of an anomaly \cite{EnergyStorage-JH-2023}. Given the importance of this role, fault diagnosis of lithium-ion battery packs has become a focus of extensive research. However, current approaches face two significant challenges: 1) they rely heavily on short-term, real-time sensor measurements, overlooking the long-term behavior of cells; and 2) many advanced fault diagnosis methods are computationally intensive, making them difficult to implement on the embedded processors typically found in BMS \cite{AppliedEnergy-RX-2020}. This paper addresses these challenges by proposing a novel approach that leverages the structural properties of battery packs to enhance fault diagnosis. Next, we will review the literature on fault diagnosis of lithium-ion battery packs. 

\subsection{Literature Review}
Existing studies on fault diagnosis can be broadly classified into two categories: model-based \cite{Arxiv-XY-2024, TPEL-SM-2021, AppliedEnergy-ZC-2016} and data-driven approaches \cite{Sensors-SM-2021, TIE-OO-2021, PowerSources-HZ-2024}. Model-based approaches construct a physical model of the battery pack and compare sensor measurements with the model's predictions, identifying any significant discrepancies as potential faults. In contrast, data-driven approaches utilize collected data to train models capable of distinguishing between normal and abnormal battery pack operation. While both approaches have their respective advantages and limitations, this paper adopts a model-based approach. Accordingly, this literature review focuses on the model-based fault diagnosis methods.

Recent research on model-based fault diagnosis for lithium-ion battery packs has grown significantly \cite{TIEMag-HX-2020, Arxiv-XY-2024}. One major group targets internal and external short circuits in lithium-ion cells \cite{IEEEAccess-HA-2024, AppliedEnergy-ZC-2016, TPEL-XY-2022, IECON-MJ-2019}. Another group focuses on diagnosing voltage and current sensor faults \cite{TPEL-XR-2019, TIM-JH-2024}, while some address thermal anomalies in pouch cells \cite{TII-ZJ-2024, TII-WP-2023}. However, running multiple fault detection schemes simultaneously in a BMS is impractical. Ideally, a single algorithm should address both cell and sensor faults. Some studies have attempted to achieve this by using structural analysis for multi-fault diagnosis \cite{TTE-CY-2022, TPEL-ZK-2022, ITEC-CY-2022}, but these methods often require numerous sensors, adding cost and complexity. Others propose active fault diagnosis \cite{TPEL-SM-2021}, though its applicability is limited to reconfigurable battery packs.

In hindsight, existing approaches fall short in providing an efficient and practical fault diagnosis solution for entire battery packs. Key challenges include: 1) the need for solutions that apply to full battery packs, not just individual cells; 2) minimizing the number of required sensors; and 3) developing algorithms capable of handling multiple fault types, including both cell and sensor faults. This paper aims to address these limitations and advance the state of the art in lithium-ion battery fault diagnosis. 

\begin{figure*}[t]
	    \centering
    \subfloat[\centering ]{{\includegraphics[trim={0cm 0cm 0cm 0cm},clip,width=10cm]{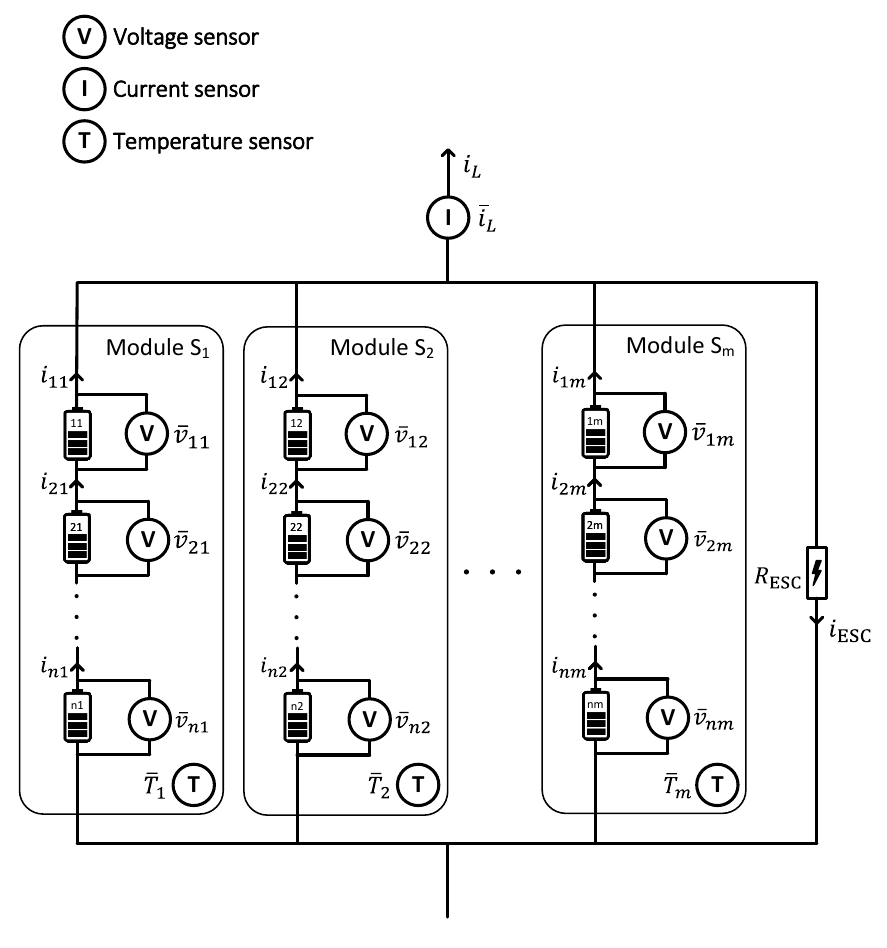} }}
	\quad
    \subfloat[\centering ]{{\includegraphics[trim={0cm 0cm 0cm 0cm},clip,width=6cm]{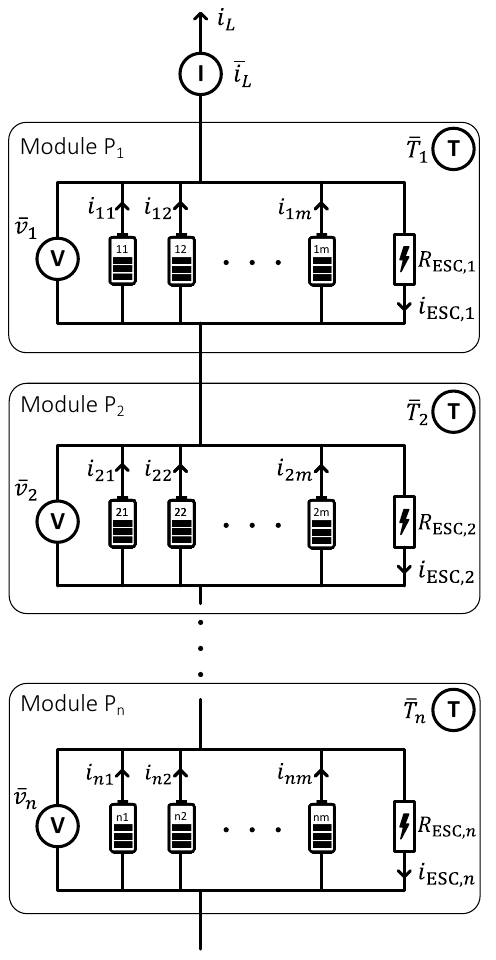} }}
    \caption{The considered battery pack configurations. (a) Series-parallel ($n$S$m$P). (b) Parallel-series ($m$P$n$S).}
    \label{FIG_1}
\end{figure*}

\subsection{Statement of Contributions}
Despite recent advances, model-based fault diagnosis approaches have yet to reach practical maturity. Existing studies typically rely solely on sensor measurements, overlooking the valuable information provided by the structural properties of the battery pack. This paper challenges this conventional approach by demonstrating how these properties can be exploited to improve fault diagnosis. The key contributions are as follows:
\begin{enumerate}[1)]
	\item We propose leveraging the structural properties of battery packs---specifically the uniformity of cells and the sparsity of fault occurrences---to enhance fault diagnosis. To achieve this, we formulate the fault diagnosis problem within the Moving Horizon Estimation (MHE) framework. The MHE problem estimates fault signals, including internal and external short circuits and current and voltage sensor faults, across various battery pack configurations (series, parallel, parallel-series, and series-parallel). This contribution enables fault diagnosis with significantly fewer sensors compared to existing studies in the literature.
	\item To address the computational complexity of MHE, we present a hierarchical solution that decomposes the large-scale, pack-level problem into smaller, module-level problems that can be solved in parallel. This approach significantly reduces computational demands while maintaining diagnostic accuracy.
\end{enumerate}
A preliminary version of this work appeared in \cite{ICPS-FA-2024}, but the current paper substantially extends the study by covering all battery pack configurations and introducing a computationally efficient hierarchical solution to the MHE problem.
\section{PROBLEM FORMULATION}
This section first presents the battery pack configurations under consideration, along with their corresponding sensor placements. Next, we introduce an electro-thermal model, and finally, we formulate the fault diagnosis problem within the MHE framework.

\subsection{Battery Pack Structure}
Fig.~\ref{FIG_1} illustrates the two battery pack configurations considered for fault diagnosis: series-parallel ($n$S$m$P) and parallel-series ($m$P$n$S). In the $n$S$m$P configuration (Fig.~\ref{FIG_1} (a)), $m$ parallel modules are formed, each consisting of $n$ serially connected cells. We equip the $n$S$m$P configuration with $n\times m$ voltage sensors, $m$ temperature sensors and a single current sensor. The voltage sensors measure the terminal voltages of all the cells and the temperature sensors measure the aggregate temperature of the cells within a module. The $m$P$n$S configuration, on the other hand, consists of $n$ serially connected modules, each containing $m$ parallel cells (Fig.~\ref{FIG_1} (b)). For this configuration,  a single current sensor measures the pack current, while each module is provided with one voltage sensor and one temperature sensor. Therefore, an $m$P$n$S configuration requires $n$ voltage and temperature sensors, along with one current sensor for the entire pack.

In both battery pack configurations, the constituent cells are sourced from the same manufacturer to minimize cell-to-cell variations within the pack. It is well-known that such variations can lead to issues like overcharging or overdischarging, which in turn may cause thermal runaway \cite{AppliedEnergy-KK-2023, AppliedEnergy-XL-2019}. Given the uniformity of the cells (with minor heterogeneities being negligible), this paper explores how this feature can be utilized for fault diagnosis. Specifically, we investigate how deviations among cells within and between modules can serve as indicators of faults. Next, we model the battery pack under fault conditions to investigate these aspects.

\subsection{Electro-thermal Modeling}
In this section, we present the electro-thermal modeling of the battery pack, considering three types of faults: 1) internal short circuit (ISC), characterized by small incremental currents through the cell separator with limited heat generation in the early stages of the fault; 2) external short circuit (ESC); and 3) voltage and current sensor faults. We also assume that all temperature sensors are functioning properly.

\begin{figure}[!t]
    \centering
    \subfloat[\centering ]{{\includegraphics[trim={0.2cm 0cm 0.2cm 0cm}, width=2.9cm]{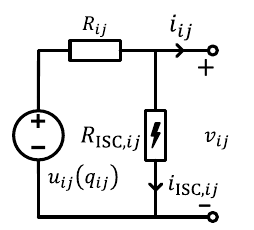} }} 
    \subfloat[\centering ]{{\includegraphics[trim={0.2cm 0cm 0.2cm 0cm}, width=5.5cm]{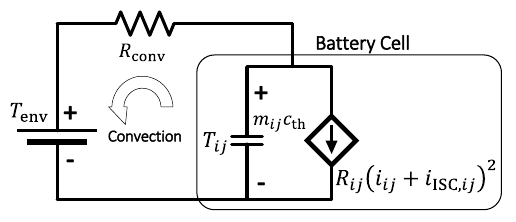} }}
    \caption{The cell-level electro-thermal model, adapted from \cite{ICPS-FA-2024}. (a) The electrical model of the cell $ij$. (b) The thermal model of the cell $ij$.}
    \label{CellLevelModel}
\end{figure}

\subsubsection{ISC fault modeling}
To model the electrical behavior of the cells, we employ the Rint model \cite{Book-PG-2015}, which consists of an open-circuit voltage (OCV) source in series with an internal resistor, as illustrated in Fig.~\ref{CellLevelModel} (a). Additionally, the ISC is represented as an extra resistor within the Rint model. The dynamics for each cell $ij$ are governed by
\begin{subequations}
	\begin{align}
		\dot{q}_{ij}(t)&=-\frac{1}{Q_{ij}}\left(i_{ij}(t) + i_{\textrm{ISC},ij}(t)\right), \label{SoCDynamics}\\
		v_{ij}(t)&=u_{ij}(q_{ij}(t))-R_{ij}\left(i_{ij}(t) + i_{\textrm{ISC},ij}(t)\right),
	\end{align}
\end{subequations}
where $q_{ij}$, $Q_{ij}$, $u_{ij}$, $v_{ij}$, and $i_{ij}$ are the cell's state-of-charge (SoC), capacity, OCV, terminal voltage, and charging/discharging current, respectively. Further, we denote the ISC resistance and current for cell $ij$ as $R_{\textrm{ISC},ij}$ and $i_{\textrm{ISC},ij}$, respectively. Without loss of generality, we also assume a linear SoC/OCV relationship as follows:
\begin{equation}
	u_{ij}(q_{ij}(t)) = \alpha + \beta q_{ij}(t),
\end{equation}
where $\alpha$ and $\beta$ are the intercept and slope coefficients.  

To model the thermal behavior of the cells, we employ a lumped capacitance model \cite{ITSE-PC-2016}. This model assumes that internal power loss is the primary source of heat generation, while heat dissipation occurs mainly through convection to the environment. The thermal dynamics for cell $ij$ are governed by
\begin{equation}
	m_{ij}c_{\textrm{th}}\dot{T}_{ij}(t) = R_{ij}\left(i_{ij}(t) + i_{\textrm{ISC},ij}(t)\right)^2 - \frac{(T_{ij}(t)-T_{\textrm{env}})}{R_{\textrm{conv}}},
	\label{CellLevelThermalModel}
\end{equation}
where $T_{ij}$ and $T_{\textrm{env}}$ are the cell's and environmental temperatures, respectively; $m_{ij}$, $c_{\textrm{th}}$, and $R_{\textrm{conv}}$ are the cell's mass, specific heat capacity, and convective thermal resistance between the cell and the environment, respectively. The electro-thermal model presented in \eqref{SoCDynamics}-\eqref{CellLevelThermalModel} is both expressive and computationally efficient, and effectively capture the cell dynamics under ISC. From \eqref{SoCDynamics} and \eqref{CellLevelThermalModel}, we see that the ISC contributes to both SoC loss and temperature increase in the cells. 

\subsubsection{ESC fault modeling}
For the $m$P$n$S configuration, we consider ESC for each module $P_i$ for $i=1,\dots,n$ (see Fig.~\ref{FIG_1} (b)). For module $P_i$, we have
\begin{equation}
	i_{i1}+i_{i2}+...+i_{im} = i_{L} + i_{\textrm{ESC},i},
	\label{PS_KCL}
\end{equation}
where $i_L$, $i_{\textrm{ESC},i}$ and $R_{\textrm{ESC},i}$ represent the battery pack current, ESC current and resistance, respectively. Conversely, for the $n$S$m$P configuration, we consider a single ESC for the entire pack, as illustrated in Fig.~\ref{FIG_1} (a). We have
\begin{equation}
	i_{11}+i_{12}+...+i_{1m} = i_{L} + i_{\textrm{ESC}}.
	\label{SP_KCL}
\end{equation}

\subsubsection{Sensor fault modeling}
For the $m$P$n$S configuration, each module $P_i$ for $i=1,\dots,n$ is equipped with a voltage sensor that measures the terminal voltages as follows:
\begin{equation}
	\bar v_{i} = v_{ij} + f_i^v, \quad j=1,\dots,m,
	\label{PS_VS}
\end{equation}
where $\bar v_{i}$ is the measured voltage for module $P_i$, and $f_i^v$ is the fault signal ($f_i^v=0$ when the sensor is functional). Also note that since the cells are in parallel in $P_i$ module, the voltages satisfy $v_{i1}=v_{i2}=\dots=v_{im}$. However, for the $n$S$m$P configuration, we measure the terminal voltages of all individual cells as follows:
\begin{equation}
	\bar v_{ij} = v_{ij} + f_{ij}^v.
	\label{SP_VS}
\end{equation}
We also measure the battery pack current in both configurations as follows:
\begin{equation}
	\bar i_L = i_L + f^i,
	\label{Isense}
\end{equation}
where $\bar i_L$ is the measured current, and $f^i$ represents the fault in the current sensor ($f^i=0$ when the sensor is functional). We monitor the cells' temperature using a temperature sensor placed in each module. In this paper, we do not account for faults for the temperature sensors and assume uniform temperature distribution within each module. For example, for the $m$P$n$S configuration, the temperature for module $P_i$ is given by
\begin{equation}
	\bar T_i = T_{i1} = T_{i2} = \dots = T_{im},
	\label{Tsense}
\end{equation}
where $\bar T_i$ is the measured temperature for module $P_i$.
\subsection{MHE Formulation}
We now proceed to formulate the MHE problem for the proposed fault diagnosis approach. The core idea is to leverage the structural properties of battery packs to enhance fault detection. Specifically, we focus on two key properties: the sparsity of fault occurrences and the uniformity among cells. First, it is highly unlikely that multiple faults will occur simultaneously across the battery pack, so it is reasonable to promote fault sparsity. Second, battery packs are composed of uniform, identical cells, which are expected to exhibit consistent behavior. Any deviation or inconsistency in the performance of these cells can be an indicator of potential faults. Here, the key question is how to exploit these properties for fault diagnosis. We propose to formulate the fault estimation as an MHE problem, where the estimation is framed as an optimization problem that allows for the incorporation of constraints \cite{Book-RC-2000}. This formulation enables us to effectively utilize the sparsity of faults and uniformity of cells within the MHE framework to enhance diagnosis.

We now formulate the MHE problem for the $m$P$n$S configuration, with the formulation for the $n$S$m$P configuration following a similar procedure. We start by discretizing the dynamic equations in \eqref{SoCDynamics} and \eqref{CellLevelThermalModel} using the forward Euler method. After discretization, to simplify the notation, we shift the discrete time index $t$ to the subscript. Further, we represent vectors with bold lowercase letters and matrices with bold uppercase letters. We then compactly express the battery pack dynamics and measurements as follows:
\begin{subequations}
	\begin{align}
		\bm{x}_{t+1} = \bm{g}\left(\bm{x}_t, \bm{u}_t, \bm{f}_t\right) + \bm{w}_t, \\
		\bm{y}_t = \bm{h}\left(\bm{x}_t, \bm{u}_t, \bm{f}_t\right) + \bm{z}_t,
	\end{align}
	\label{Dynamics}
\end{subequations}
where 
\begin{align}
	\bm x =& \begin{bmatrix}\bm x_1^\top & \dots & \bm x_n^\top \end{bmatrix}^\top, \nonumber \\
	\bm u =& \begin{bmatrix}\bm u_1^\top & \dots & \bm u_n^\top \end{bmatrix}^\top, \nonumber \\
	\bm f =& \begin{bmatrix}\bm f_{\textrm{ISC}}^\top & \bm f_{\textrm{ESC}}^\top & \bm f_{v,i}^\top \end{bmatrix}^\top, \nonumber \\
	\bm y =& \begin{bmatrix}\bm y_1^\top & \dots & \bm y_n^\top & \bar i_L \end{bmatrix}^\top, \nonumber
\end{align}
with 
\begin{align}
	\bm{x}_i &= \begin{bmatrix}q_{i1}\\\vdots\\q_{im}\\T_{i1}\\\vdots\\T_{im}\end{bmatrix}, \;
	\bm{u}_i = \begin{bmatrix}i_{i1}\\\vdots\\i_{im}\end{bmatrix}, \;
	\bm{y}_i = \begin{bmatrix}\bar v_i \\ \bar T_i\end{bmatrix}, \; \nonumber \\ 
	\bm{f}_{\textrm{ISC}} &= \begin{bmatrix}i_{\textrm{ISC},11}\\ \vdots \\i_{\textrm{ISC},nm}\end{bmatrix}, \;
	\bm{f}_{\textrm{ESC}} = \begin{bmatrix}i_{\textrm{ESC},1}\\ \vdots \\i_{\textrm{ESC},n} \end{bmatrix}, \; 
	\bm{f}_{v,i} = \begin{bmatrix}f_1^v \\\vdots\\f_n^v \\f^i \end{bmatrix}, \; \nonumber 
\end{align}
for $i=1,\dots,n$. Above, the function $\bm g$ contains \eqref{SoCDynamics} and \eqref{CellLevelThermalModel} for all cells, while $\bm h$ collects the measurements in \eqref{PS_VS}-\eqref{Tsense}. Further, $\bm w_t\in\mathbb{R}^{2mn}$ and $\bm z_t\in\mathbb{R}^{2n+1}$ denote the bounded process and measurement disturbances, respectively. 

Moving forward, we define the cost function for the MHE problem. As previously discussed, we aim to promote the sparsity of fault occurrences. To achieve this, we penalize the $\ell_0$-norm of the incremental changes in the fault signals within the cost function. The cost function is formulated as follows:
\begin{equation}
	\begin{split}
		\phi_t = \sum_{t=k-H}^{k-1} &\norm{\bm{w}_t}_{\bm{Q}}^2 + \norm{\bm{z}_t}_{\bm{R}}^2 + \norm{\Delta \bm{f_t}}_0  \\
			& + \norm{\bm{z}_k}_{\bm{R}}^2 + \phi_{\textrm{ac}}\left(\hat{\bm{x}}_{k-H}\right),
		\end{split}
	\label{CostFunction-1}
\end{equation}
where $\phi_t$ represents the cost function at time index $t$, $H$ is the horizon length, and $\bm Q$ and $\bm R$ are weight matrices. The term $\norm{(\cdot)}_{\bm Q}^2$ is defined as $(\cdot)^\top\bm Q(\cdot)$, while $\norm{(\cdot)}_0$ denotes the $\ell_0$-norm operator. Further, $\Delta \bm f_t$ is the incremental changes in fault signals given by  
\begin{equation}
	\Delta \bm{f}_t = \bm{f}_t - \bm{f}_{t-1}.
\end{equation}
Note that $\hat{\bm{x}}_{k-H}$ is the estimate of $\bm x$ at time $k-H$, and $\phi_{\textrm{ac}}(\hat{\bm{x}}_{k-H})$ denotes the arrival cost. The arrival cost incorporates prior information from time $t=0$ to $t=k-H$, beyond the estimation horizon. While the arrival cost significantly impacts the performance of MHE, its detailed computation is beyond the scope of this study. However, several studies provide methods for its calculation \cite{IFAC-XL-2024, TAC-BG-2019}.

We have addressed the sparsity of fault occurrences by applying $\ell_0$-norm penalization to the incremental changes in fault signals. To promote uniformity among the cells, we incorporate constraints. Specifically, we impose limits on the allowable variations in the cells' SoC and temperature values as follows:
\begin{subequations}
	\begin{align}
		&\left| q_{ij} - q_{i'j'} \right| \leq \Delta q,\\
		&\left| T_{ij} - T_{i'j'} \right| \leq \Delta T,
	\end{align}
	\label{Uniformity}
\end{subequations}
\hspace{-5pt}where $\Delta q$ and $\Delta T$ are the maximum allowable deviation in cells' SoC and temperature values, respectively. It is important to note that the constraints in \eqref{Uniformity} are applied across all possible cell pairs ($ij$, $i'j'$) within the battery pack. By limiting the discrepancies between cells during normal operation, any violations of these constraints are treated as potential indicators of faults.

We also introduce lower and upper bounds on the fault signals and disturbances as follows:
\begin{subequations}
	\begin{align}
		& \bm{f}_t^{\textrm{min}}\leq \bm{f}_t\leq \bm{f}_t^{\textrm{max}},\\
		& \bm{w}_t^{\textrm{min}}\leq \bm{w}_t\leq \bm{w}_t^{\textrm{max}},\\
		& \bm{z}_t^{\textrm{min}}\leq \bm{z}_t\leq \bm{z}_t^{\textrm{max}},
	\end{align}
\end{subequations}
where $\bm{f}_t^{\textrm{min}/\textrm{max}}$, $\bm{w}_t^{\textrm{min}/\textrm{max}}$, and $\bm{z}_t^{\textrm{min}/\textrm{max}}$ are the lower/upper bounds on faults and disturbances, respectively. These constraints enable the proposed approach to distinguish between different fault types, such as ISC and ESC, as well as to differentiate disturbances from actual fault signals. 

Having laid out the cost function and the constraints, we are now ready to formulate the MHE problem. It is
\begin{equation}
	\begin{aligned}
		&\min_{\bm w_t, \bm z_t, \bm f_t, \hat{\bm{x}}_{k-H}} \quad \sum_{t=k-H}^{k-1} \norm{\bm{w}_t}_{\bm{Q}}^2 + \norm{\bm{z}_t}_{\bm{R}}^2 + \norm{\Delta \bm{f_t}}_0 \\
		& \qquad \qquad \qquad \qquad + \norm{\bm{z}_k}_{\bm{R}}^2 + \phi_{\textrm{ac}}\left(\hat{\bm{x}}_{k-H}\right),\\
		&\textrm{Cell dynamics:} \quad \\
		&\ \bm{x}_{t+1} = \bm{g}\left(\bm{x}_t, \bm{u}_t, \bm{f}_t\right) + \bm{w}_t, \\
		&\ \bm{y}_t = \bm{h}\left(\bm{x}_t, \bm{u}_t, \bm{f}_t\right) + \bm{z}_t,\\
		&\textrm{Incremental fault dynamics:} \quad \\
		&\ \Delta \bm{f}_t = \bm{f}_t - \bm{f}_{t-1}, \\
		&\textrm{Fault and noise constraints:} \quad \\
		&\ \bm{f}_t^{\textrm{min}}\leq \bm{f}_t\leq \bm{f}_t^{\textrm{max}}, \\ 
		&\ \bm{w}_t^{\textrm{min}}\leq \bm{w}_t\leq \bm{w}_t^{\textrm{max}},\\
		&\ \bm{z}_t^{\textrm{min}}\leq \bm{z}_t\leq \bm{z}_t^{\textrm{max}},\\
		&\textrm{Uniformity constraints:} \quad \\
		&\ \left| q_{ij} - q_{i'j'} \right| \leq \Delta q,\\
		&\ \left| T_{ij} - T_{i'j'} \right| \leq \Delta T,
	\end{aligned}
	\label{MHE}
\end{equation}
where $\Delta t$ is the sampling time. The problem formulated in \eqref{MHE} effectively leverages both the sparsity of fault occurrences and the uniformity among cells to estimate fault signals. However, two significant challenges arise. First, the inclusion of the $\ell_0$-norm renders the problem nonlinear and nonconvex, making it difficult to obtain a tractable solution \cite{Optimization-ZD-2023}. Second, the number of optimization variables and constraints increases with the number of cells, making it computationally feasible for small-scale battery packs but prohibitive for larger-scale ones. To address these limitations, we propose a hierarchical solution to \eqref{MHE} in the following section.
\section{THE PROPOSED HIERARCHICAL APPROACH}
Although the problem in \eqref{MHE} leverages structural properties to improve fault diagnosis, solving it for the entire battery pack is computationally formidable. The bottleneck arises from two factors: 1) the use of $\ell_0$-norm regularization in \eqref{CostFunction-1} and 2) the large number of constituent cells. First, we address the $\ell_0$-norm regularization in \eqref{CostFunction-1} by relaxing it with a mixed $\ell_{2,1}$-norm, as follows:
\begin{equation}
	\norm{\Delta \bm{f_t}}_{2,1} = \norm{\begin{bmatrix} \lambda_{\textrm{ISC}}\norm{\Delta \bm f_{\textrm{ISC}}}_2 \\ \lambda_{\textrm{ESC}}\norm{\Delta \bm f_{\textrm{ESC}}}_2 \\ \lambda_{v,i}\norm{\Delta \bm f_{v,i}}_2\end{bmatrix}}_1
\end{equation}
where $\lambda_{\textrm{ISC}}$, $\lambda_{\textrm{ESC}}$, and $\lambda_{v,i}$ are the respective weights for each fault type, and $\norm{\cdot}_1$ and $\norm{\cdot}_2$ denotes the $\ell_1$-norm and $\ell_2$-norm, respectively. 

To tackle the computational burden from the large number of cells, we propose a hierarchical approach. The key idea is to first detect the fault's location at the module level, and subsequently perform a more detailed cell-level diagnosis within the suspected faulty module. This process begins with an inter-module diagnosis to identify the suspected module and is followed by intra-module diagnoses to pinpoint the specific faulty cell within that module. Note that this approach leverages both the uniformity of the cells (modules) and sparsity of faults in both the inter-module  and intra-module problems. The proposed hierarchical approach comprises three steps, outlined as follows.

\begin{figure}[!t]
    \centering
    \subfloat[\centering ]{{\includegraphics[trim={0.2cm 0cm 0.2cm 0cm}, width=7cm]{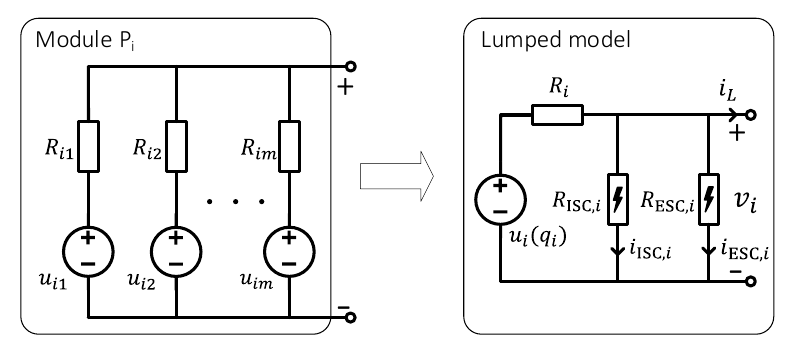} }} \\
    \subfloat[\centering ]{{\includegraphics[trim={0.2cm 0cm 0.2cm 0cm}, width=7cm]{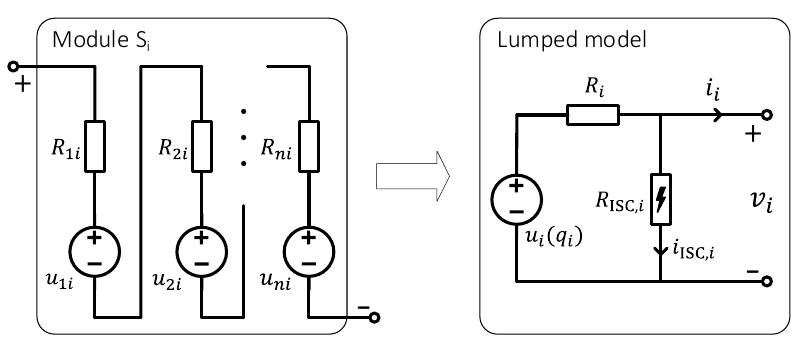} }}
    \caption{The lumped module-level electro-thermal model. (a) Lumped model for module $P_i$. (b) Lumped model for module $S_i$.}
    \label{AggregateModel}
\end{figure}

{\it Step 1: Lumped module modeling.} To perform inter-module fault diagnosis, we need an electro-thermal model for each module. This step aggregates the cells within each module and represents them with a simplified lumped model \cite{TTE-FA-2024}. Let us consider module $P_i$ in the $m$P$n$S configuration, which consists of $m$ cells in parallel. The lumped model for these cells, illustrated in Fig.~\ref{AggregateModel} (a), captures the module's electrical dynamics as follows:
\begin{equation}
	Q_{i} = \sum_{j=1}^{m} Q_{ij}, \quad \dot q_i(t) = -\frac{1}{Q_i}(i_L+i_{\textrm{ISC},i}+i_{\textrm{ESC},i}),
\end{equation}
where $Q_i$ and $q_i$ represent the aggregated capacity and SoC of module $P_i$, respectively. Additionally, the aggregate internal resistance and OCV are given by:
\begin{equation}
	R_i = \frac{1}{\sum_{j=1}^{m} \frac{1}{R_{ij}}}, \qquad u_i = R_i \sum_{j=1}^{m}\frac{u_{ij}}{R_{ij}},
\end{equation}
where $R_i$ and $u_i$ denote the module's overall internal resistance and OCV, respectively. We also express the thermal dynamics of the lumped model as follows:
\begin{equation}
	m_{i}c_{\textrm{th}}\dot{T}_{i}(t) = R_{i}\left(i_{L} + i_{\textrm{ISC},i} + i_{\textrm{ESC},i}\right)^2 - \frac{(T_{i}(t)-T_{\textrm{env}})}{R_{\textrm{conv},i}},
	\label{PS-LumpedThermal}
\end{equation}
where $m_i$ is the total mass of module $P_i$, and $R_{\textrm{conv},i}$ is its convective heat resistance. Similarly, we develop a lumped model for module $S_i$ in the $n$S$m$P configuration as depicted in Fig.~\ref{AggregateModel} (b). The lumped model's dynamics are governed by 
\begin{subequations}
	\begin{align}
		Q_{i} &= \min\{Q_{1i}, \dots, Q_{ni}\}, \\ 
		\dot q_i(t) &= -\frac{1}{Q_i}(i_i+i_{\textrm{ISC},i}),\\
		R_i &= \sum_{j=1}^{n} R_{ji}, \quad u_i = \sum_{j=1}^{n} u_{ji}, \\
		m_{i}c_{\textrm{th}}\dot{T}_{i}(t) &= R_{i}\left(i_{i} + i_{\textrm{ISC},i}\right)^2 - \frac{(T_{i}(t)-T_{\textrm{env}})}{R_{\textrm{conv},i}},
	\end{align}
\end{subequations}
where $i_i=i_{1i}=\dots=i_{ni}$. At its core, this step simplifies the $m$P$n$S and $n$S$m$P configurations to their equivalent $1$P$n$S and $1$S$m$P configurations, respectively. This reduction in complexity significantly lowers the number of cells to be considered in \eqref{MHE}, and paves the path for the inter-module fault diagnosis in subsequent steps.

{\it Step 2: Inter-module fault diagnosis.} In this step, we aim to identify suspected faulty modules before conducting a more in-depth intra-module fault diagnosis. To do so, we formulate an inter-module fault diagnosis problem using lumped module models, similar to \eqref{MHE}. By leveraging the uniformity among modules and the sparsity of fault occurrences, this approach identifies which module may contain a fault. Consequently, it guides a more targeted intra-module fault diagnosis. The inter-module diagnosis problem is computationally efficient because it focuses on modules rather than individual cells.

{\it Step 3: Intra-module fault diagnosis.} After the inter-module diagnosis identifies faulty modules, we proceed with intra-module diagnosis. This inter-module problem mirrors the problem in \eqref{MHE} and only deals with a specific module. This step enhances computational efficiency for two reasons. First, we can run the intra-module problems in parallel. Second, because they pertain only to individual modules, they involve significantly fewer optimization variables and constraints compared to the pack-level problem in \eqref{MHE}. Overall, the proposed hierarchical approach decomposes the pack-level problem in \eqref{MHE} into a single inter-module and several intra-module problems, significantly lowering the computational requirements of fault detection for a battery pack.

\section{SIMULATION RESULTS}
   
\begin{table}[!t]
	\renewcommand{\arraystretch}{1.2}
	\caption{Specifications of the considered battery pack}
	\centering
	\label{Table_1}
	\resizebox{\columnwidth}{!}{
		\begin{tabular}{l l l}
			\hline\hline \\[-3mm]
			\multicolumn{1}{c}{Symbol} & \multicolumn{1}{c}{Parameter} & \multicolumn{1}{c}{Value [Unit]}  \\[1.6ex] 		                   \hline
			$ v $  & Cell nominal voltage & 3.6     [V] \\
			$ Q $ & Cell nominal capacity & 2.5     [Ah] \\ 
			$ R $ & Cell internal resistance & 31.3     [m$\Omega$] \\
			$ C_{\textrm{th}} $ & Thermal capacitance & 40.23     [J/K] \\ 
			$ R_{\textrm{conv}} $ & Convection thermal resistance & 41.05     [K/W] \\ 
			$ T_{\textrm{env}} $ & Environment temperature & 298     [K] \\ 
			$ \Delta q $ & SoC deviation threshold & 0.5\% \\
			$ \Delta T $ & Temperature deviation threshold & 0.5    [K] \\
			$ \Delta t $ & Sampling time & 30     [s]\\
			$ H $ & Horizon length & 300     [s]\\
			\hline\hline
		\end{tabular}
	}
\end{table}

In this section, we perform simulations across various battery pack configurations and fault types to assess the performance of the proposed approach. Specifically, we analyze two battery pack configurations: $3$P$2$S and $3$S$2$P. These battery packs consists of six Samsung INR18650-25R cells, with specifications summarized in Table~\ref{Table_1}. We initialize the simulations with the cells having SoC values of 0.9 and discharge both battery packs at a constant rate of 6 A. For both the inter-module and intra-module MHE problems, we set a sampling time and horizon length of 30 and 300 seconds, respectively. We also use the fmincon optimization package in the Matlab software to solve the MHE problems.

\subsection{$3$P$2$S battery pack} 

\begin{figure*}[t]
	\centering
    	\subfloat[\centering ]{{\includegraphics[trim={0.5cm 0 1.2cm 0cm},clip,width=5cm]{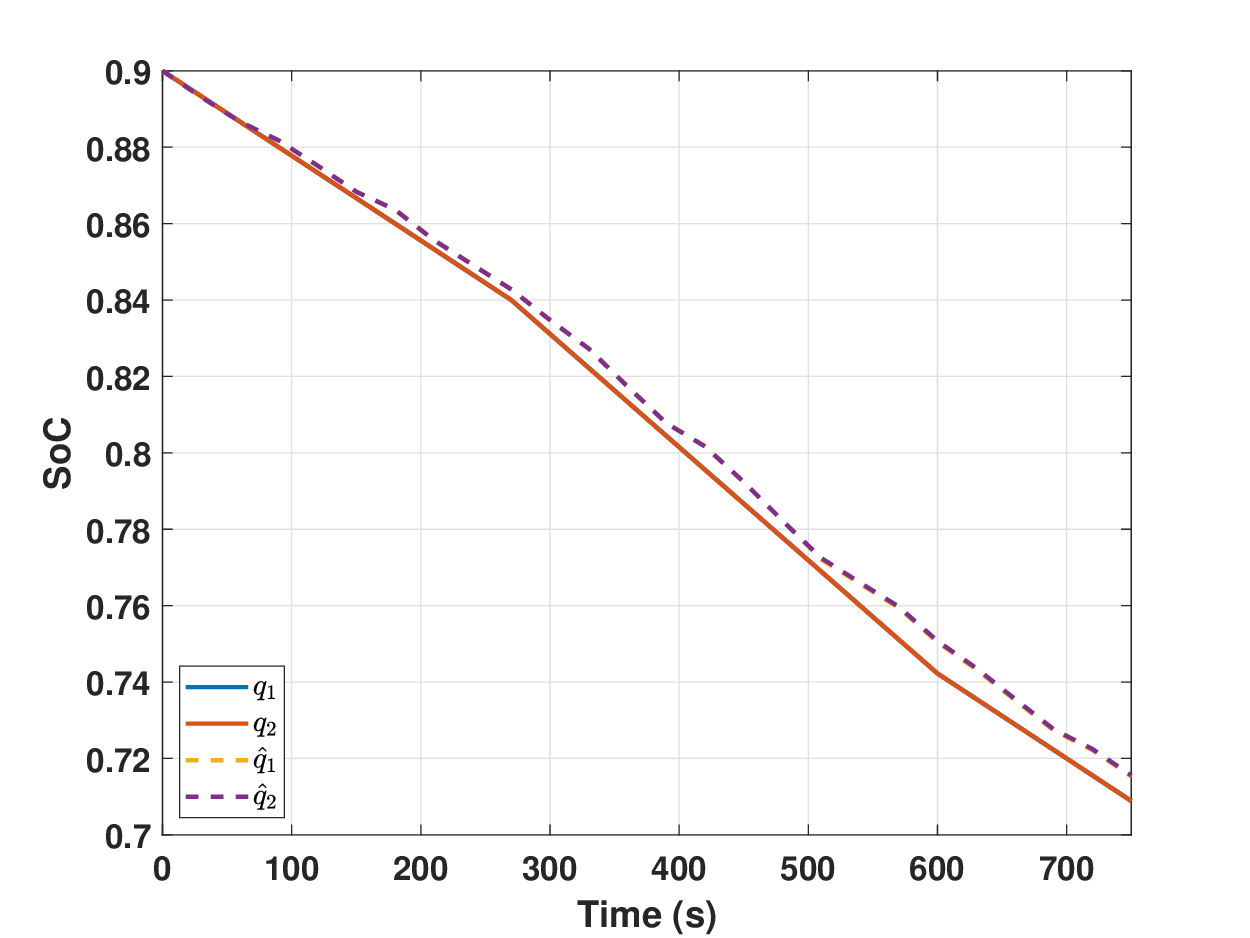} }}
	\,
    	\subfloat[\centering ]{{\includegraphics[trim={0.5cm 0 1.2cm 0cm},clip,width=5cm]{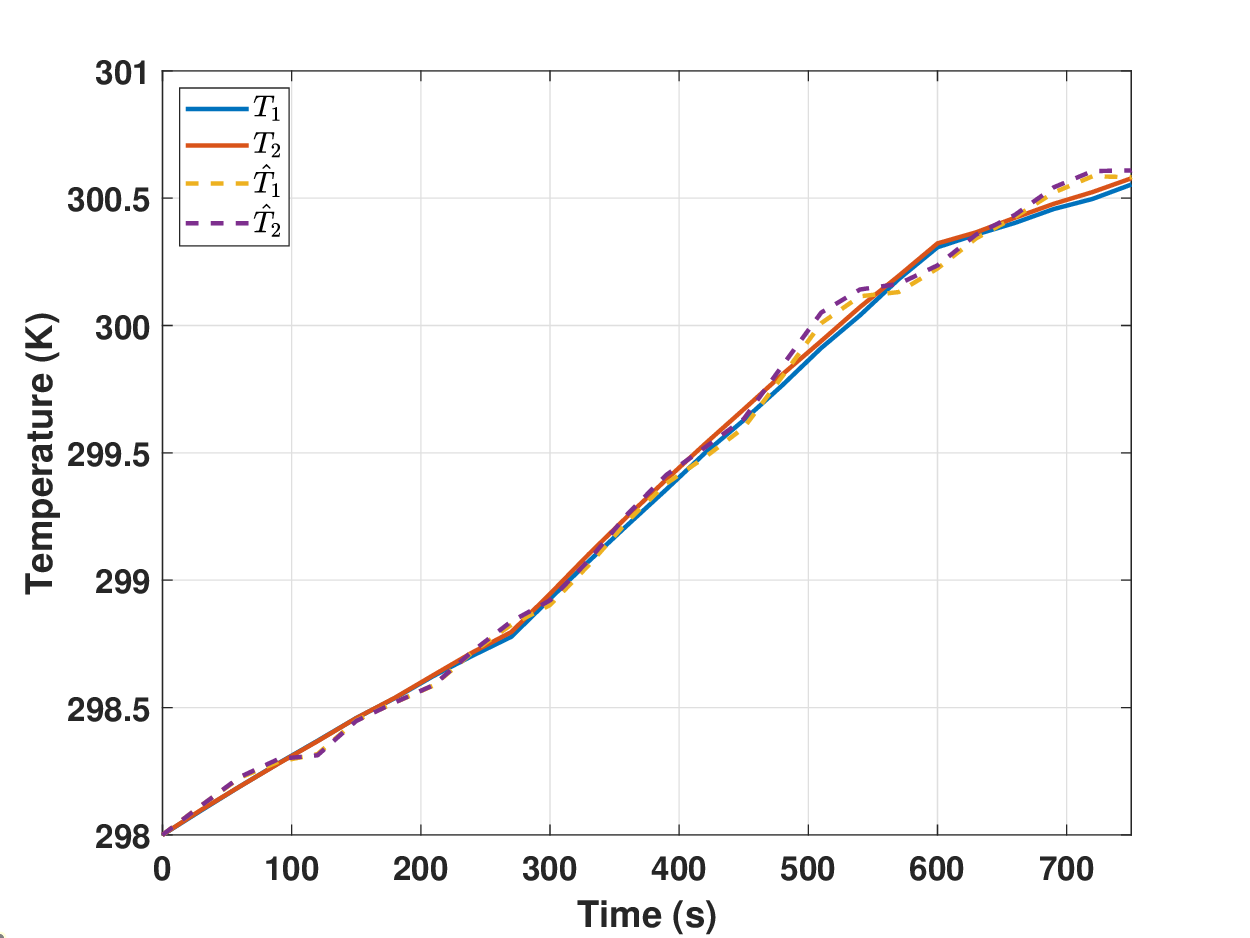} }}
    	\,
    	\subfloat[\centering ]{{\includegraphics[trim={0.5cm 0 1.2cm 0cm},clip,width=5cm]{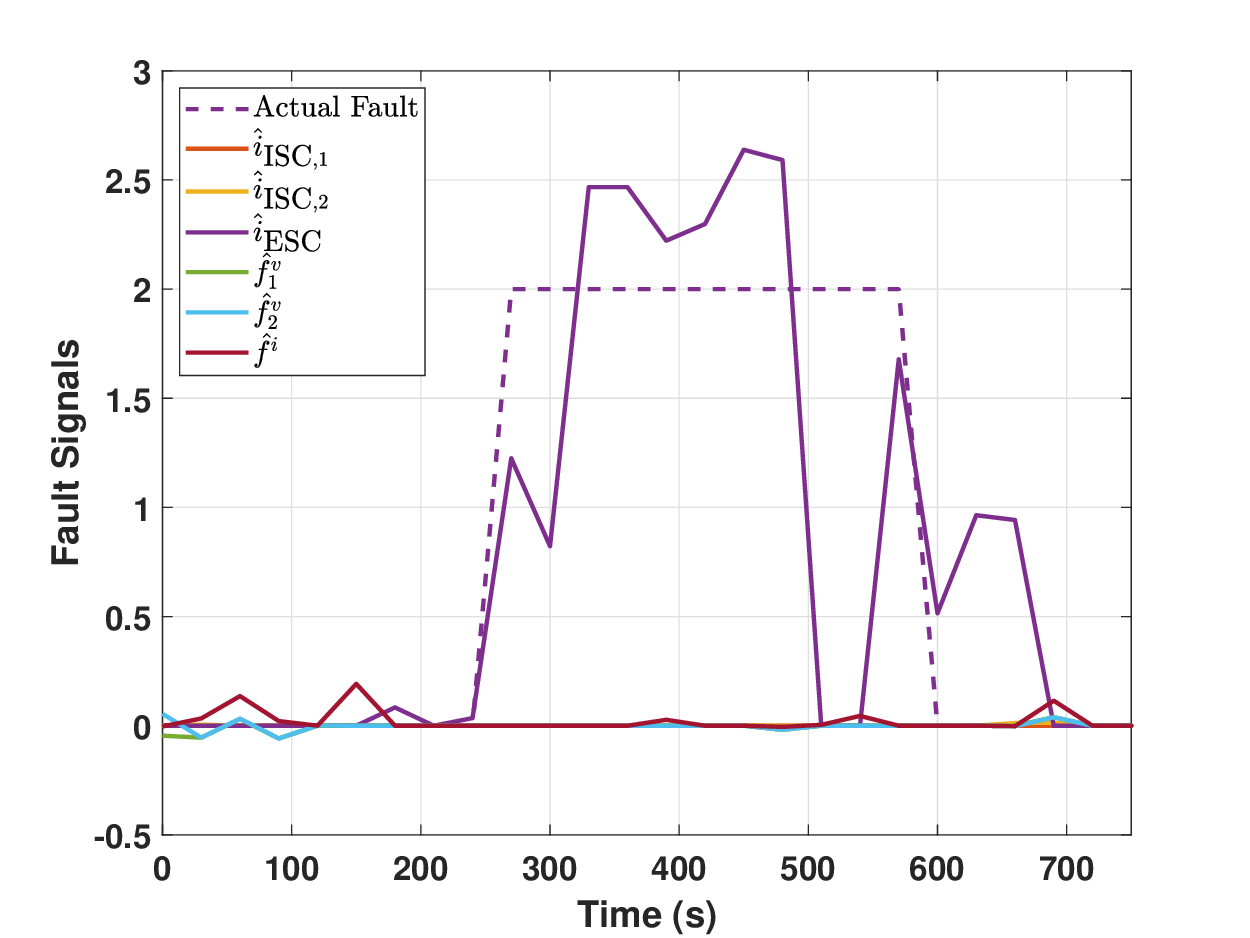} }}
    	\,    	
    	\subfloat[\centering ]{{\includegraphics[trim={0.5cm 0 1.2cm 0cm},clip,width=5cm]{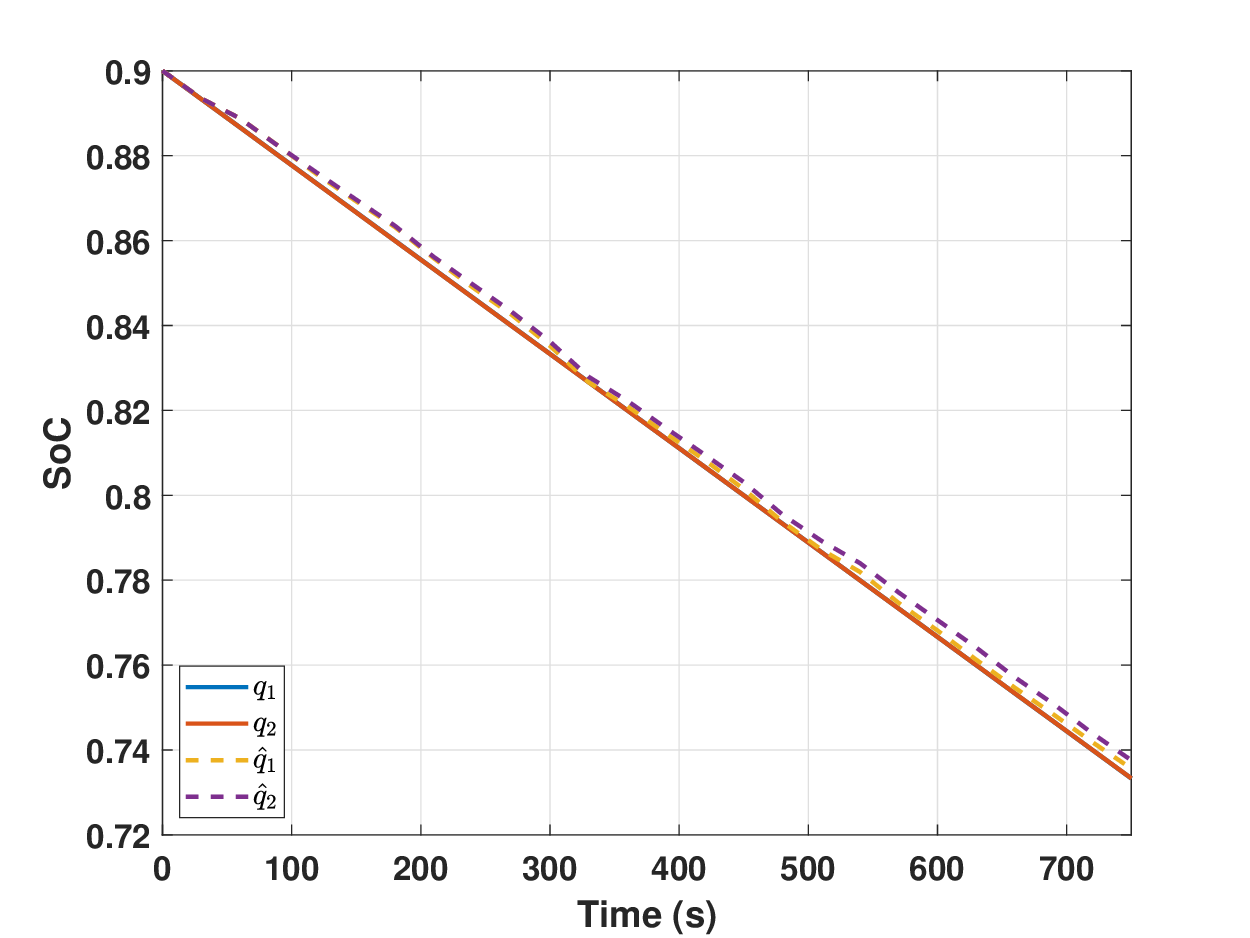} }}
	\,
    	\subfloat[\centering ]{{\includegraphics[trim={0.5cm 0 1.2cm 0cm},clip,width=5cm]{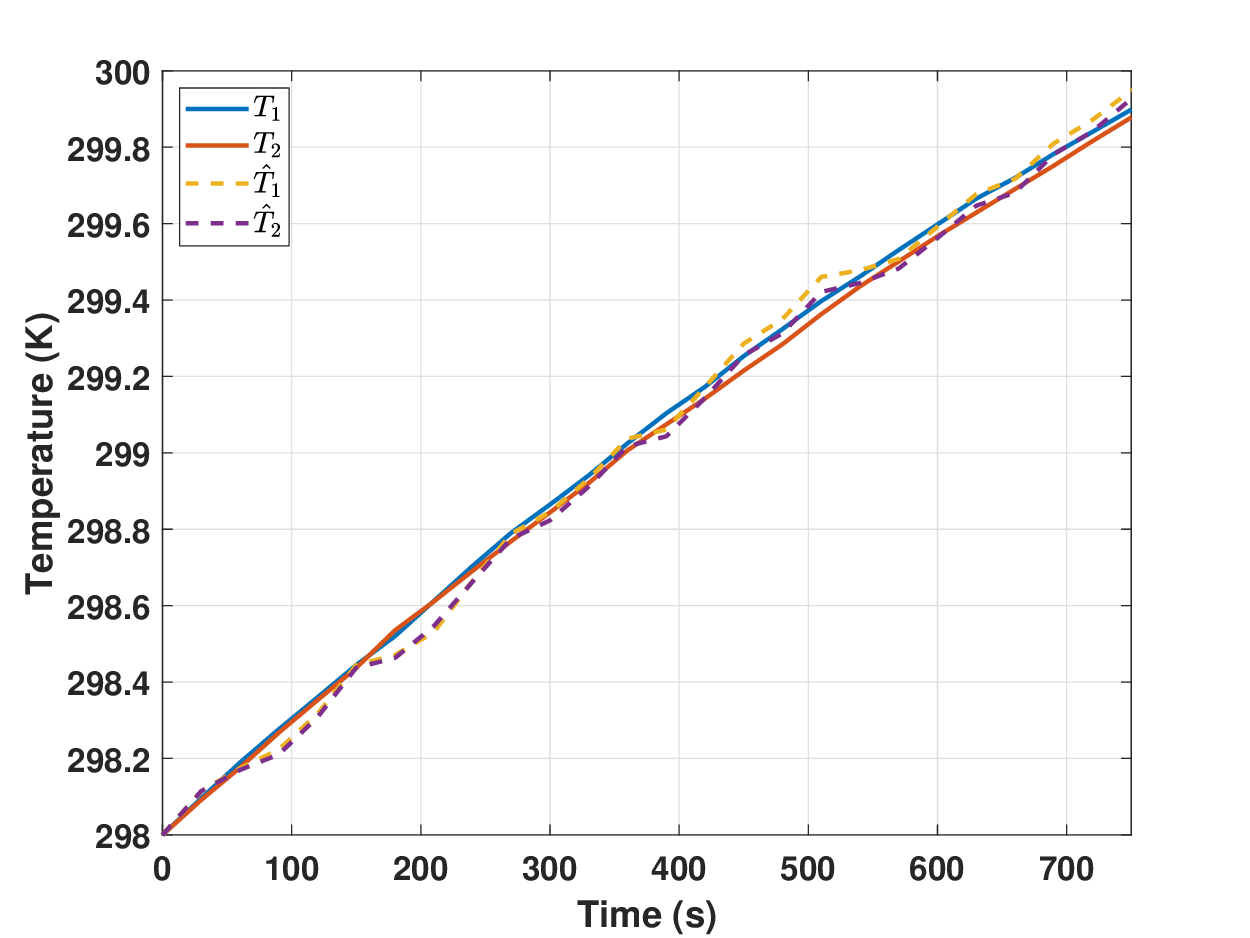} }}
    	\,
    	\subfloat[\centering ]{{\includegraphics[trim={0.5cm 0 1.2cm 0cm},clip,width=5cm]{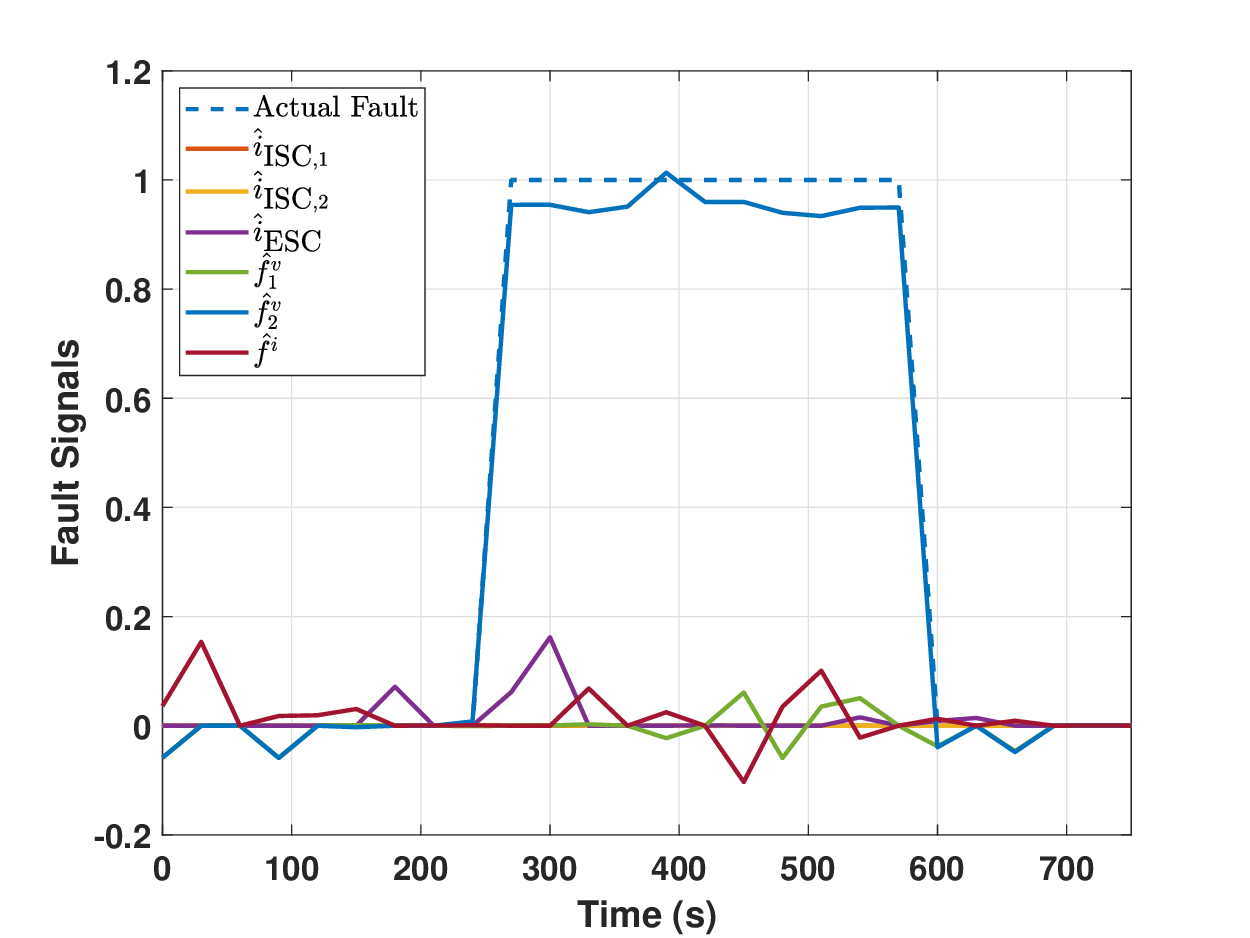} }}

    \caption{Simulation results of the inter-module diagnosis for the $3$P$2$S configuration. (a) The modules' SoC under ESC. (b) The modules' temperature under ESC. (c) The estimated faults under ESC. (d) The modules' SoC under voltage sensor fault. (e) The modules' temperature under voltage sensor fault. (f) The estimated faults under voltage sensor fault.}
    \label{FIG_SIM_1}
\end{figure*}

In this simulation, we begin with a fault-free system. At $t=250$ seconds, we introduce an ESC in module $P_1$, characterized by $i_{\textrm{ESC},1} = 2$ A. This fault is later cleared at time instant $t=600$ seconds. We first execute the inter-module problem for modules $P_1$ and $P_2$. Figs.~\ref{FIG_SIM_1} (a)-(c) depict the obtained results of the inter-module problem. As shown in Figs.~\ref{FIG_SIM_1} (a)-(b), the ESC leads to a noticeable drop in the SoC and a rise in temperature within the modules. Importantly, Fig.~\ref{FIG_SIM_1} (c) shows that the proposed approach successfully detects the ESC and accurately estimates its magnitude using only the inter-module diagnosis. 

Next, we induce a voltage sensor fault in module $P_2$ with $f_2^v = 1$ V. Having a closer look at Figs.~\ref{FIG_SIM_1} (d)-(e), we see that the proposed approach accurately estimates the modules' SoC and temperature values. This is because of the successful estimation of the voltage sensor fault signal, as illustrated in Fig.~\ref{FIG_SIM_1} (f).

For both the ESC and voltage sensor fault, we also run the intra-module problems, which yield results consistent with those shown in Figure~\ref{FIG_SIM_1}. It is worth mentioning that if the inter-module problem is successful at estimating the fault signals, there is no further need of running the intra-module problems. This is possible because the inter-module problem effectively utilizes the uniformity between modules $P_1$ and $P_2$, as well as the sparsity of fault occurrences in the diagnosis process.

\subsection{$3$S$2$P battery pack}

\begin{figure*}[t]
	\centering
    	\subfloat[\centering ]{{\includegraphics[trim={0.5cm 0 1.2cm 0cm},clip,width=5cm]{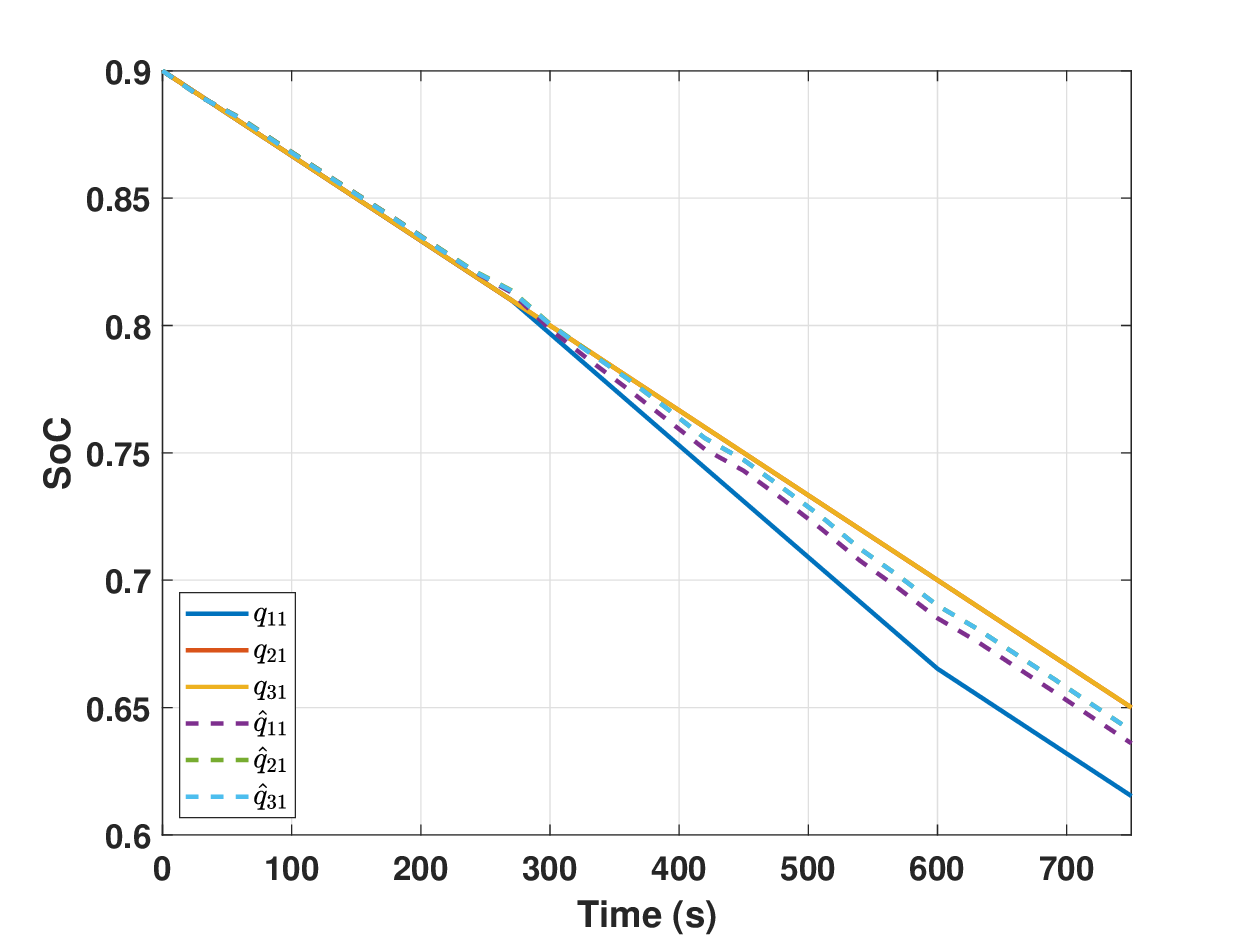} }}
	\,
    	\subfloat[\centering ]{{\includegraphics[trim={0.5cm 0 1.2cm 0cm},clip,width=5cm]{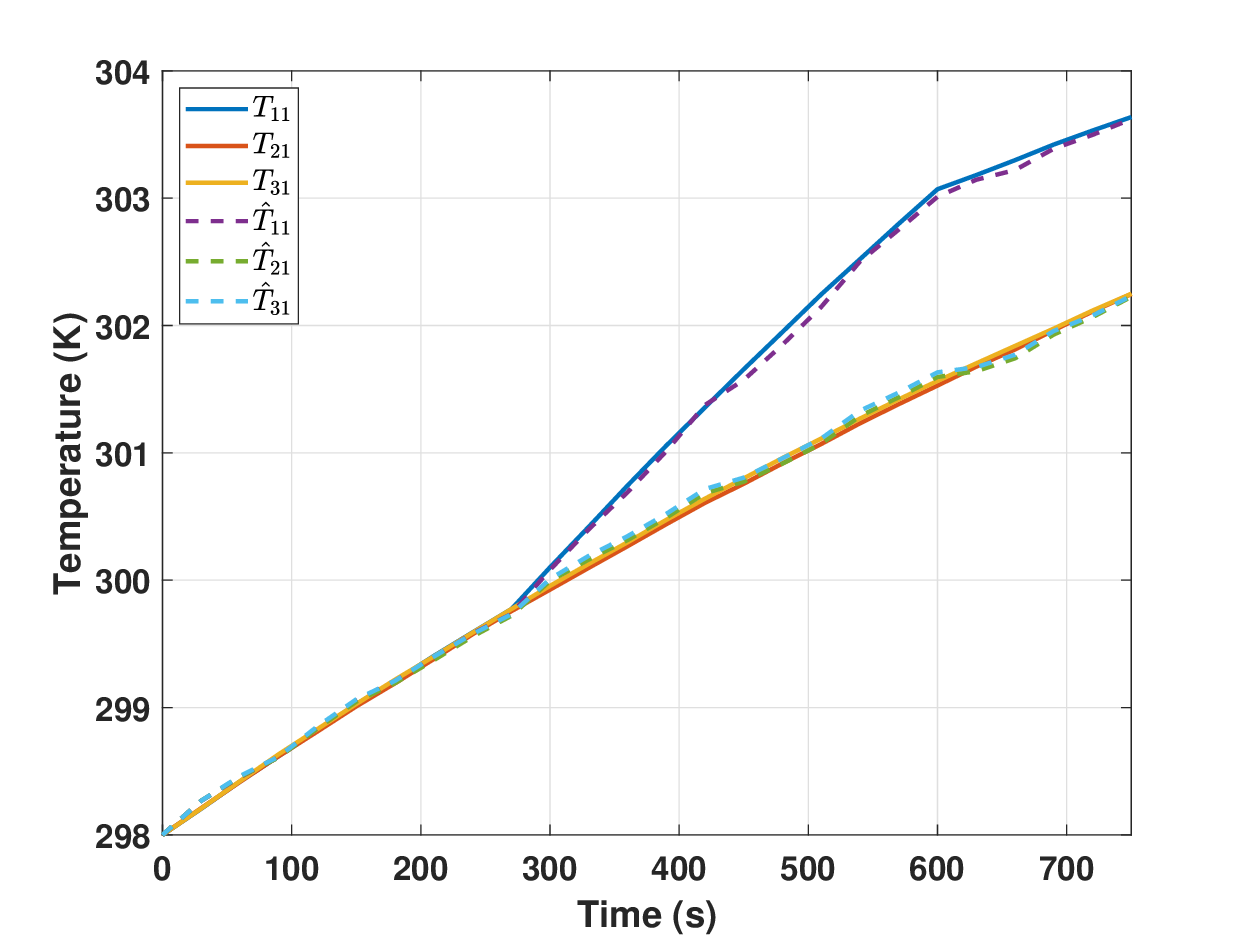} }}
    	\,
    	\subfloat[\centering ]{{\includegraphics[trim={0.5cm 0 1.2cm 0cm},clip,width=5cm]{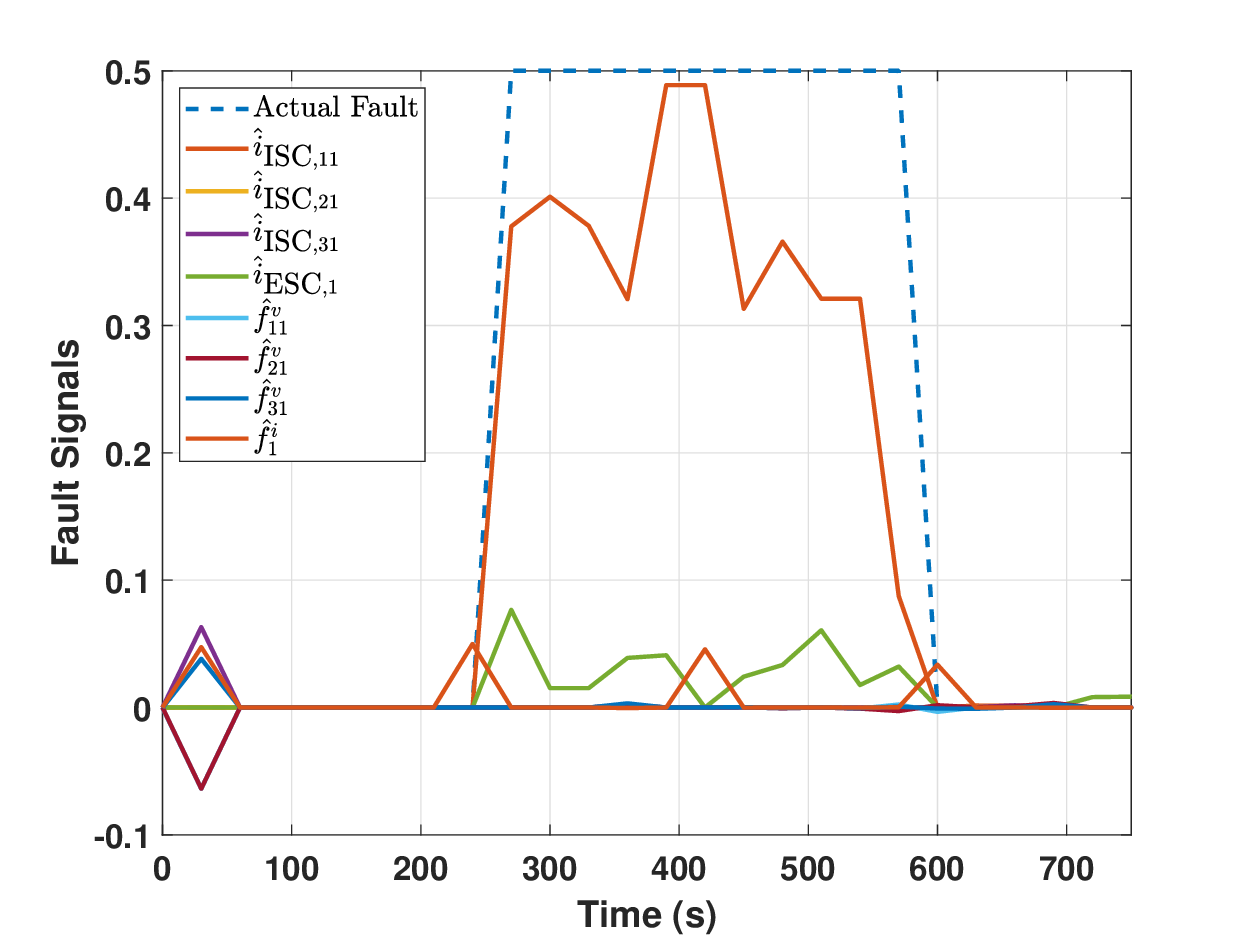} }}
    	\,    	
    	\subfloat[\centering ]{{\includegraphics[trim={0.5cm 0 1.2cm 0cm},clip,width=5cm]{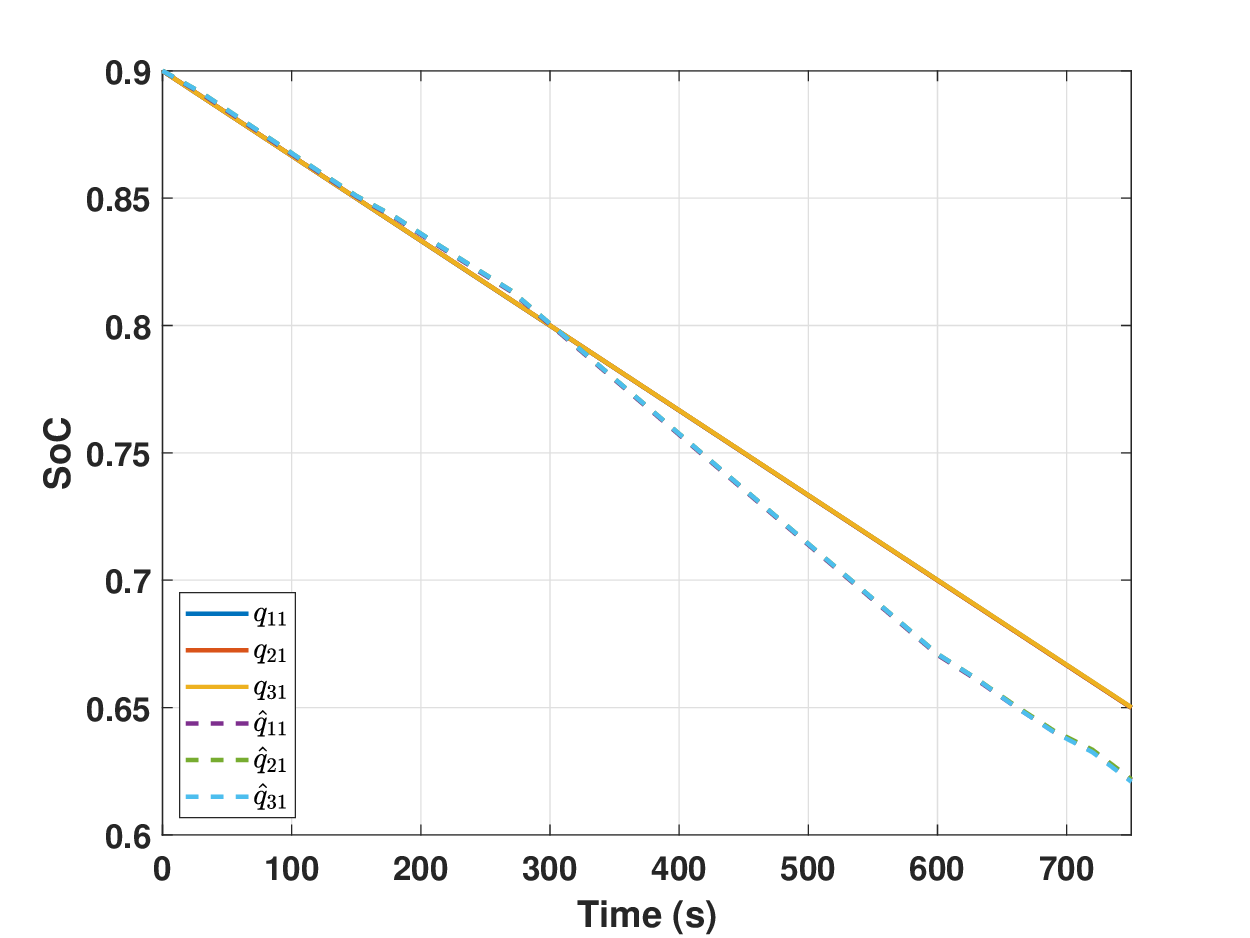} }}
	\,
    	\subfloat[\centering ]{{\includegraphics[trim={0.5cm 0 1.2cm 0cm},clip,width=5cm]{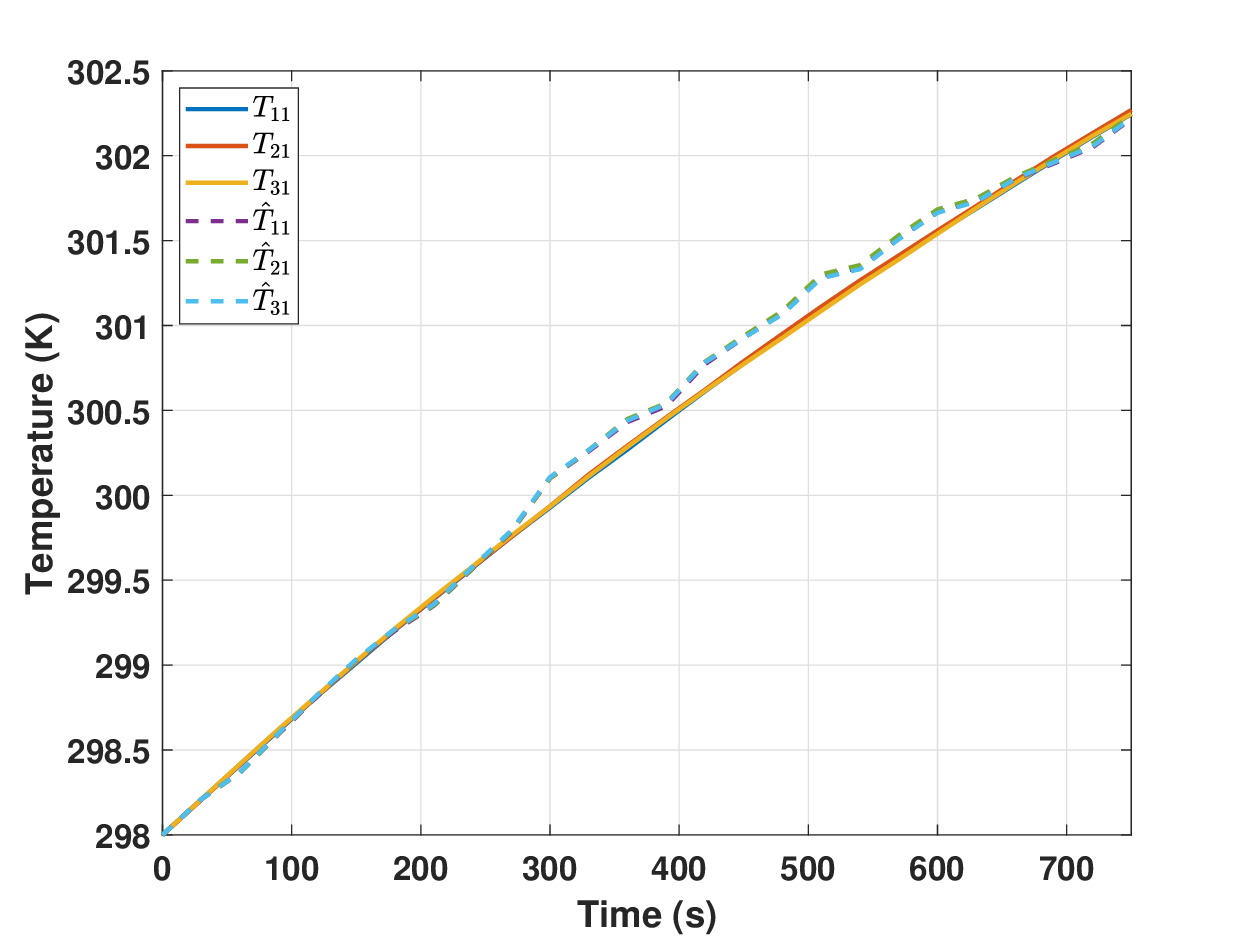} }}
    	\,
    	\subfloat[\centering ]{{\includegraphics[trim={0.5cm 0 1.2cm 0cm},clip,width=5cm]{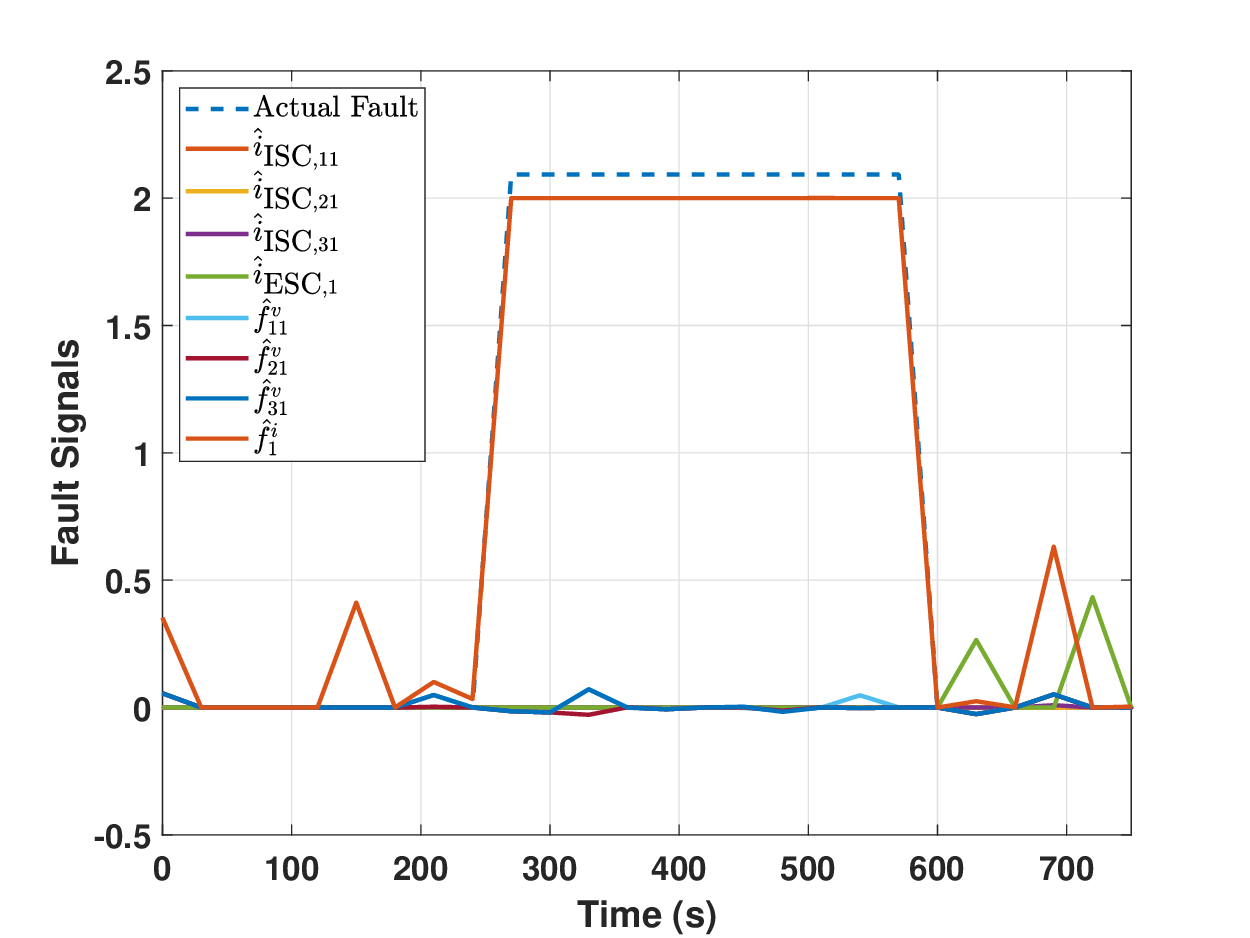} }}

    \caption{Simulation results of the intra-module diagnosis for the $3$S$2$P configuration. (a) The cell's SoC under ISC. (b) The cells' temperature under ISC. (c) The estimated faults under ISC. (d) The cells' SoC under current sensor fault. (e) The cells' temperature under current sensor fault. (f) The estimated faults under current sensor fault.}
    \label{FIG_SIM_2}
\end{figure*}

For this battery pack configuration, we introduce an ISC in cell 11, located within module $S_1$. We provide the results of the intra-module fault diagnosis for module $S_1$ in Figs.~\ref{FIG_SIM_2} (a)-(c). Figs.~\ref{FIG_SIM_2} (a)-(b) present the SoC and temperature values of the three cells in module $S_1$, where we observe a significant reduction in the SoC of cell 11 compared to the other cells, as well as a higher temperature. The intra-module diagnosis leverages these discrepancies in SoC and temperature to detect the ISC. Fig.~\ref{FIG_SIM_2} (c) demonstrates that the intra-module diagnosis successfully detects and localizes the ISC in cell 11. Additionally, we introduce a current sensor fault with $f^i=2$ A. Figs.~\ref{FIG_SIM_2} (d)-(f) report the intra-module diagnosis results, which similarly identifies the current sensor fault with high accuracy.

\section{CONCLUSIONS}
As lithium-ion battery systems become the dominant choice for energy storage, ensuring their safe and reliable operation is increasingly critical. Fault diagnosis plays a key role in achieving this goal. This paper presents an enhanced fault diagnosis approach for lithium-ion battery packs, leveraging key structural properties such as the uniformity among constituent cells and the sparsity of fault occurrences. The proposed fault diagnosis specifically targets internal and external short circuits, as well as voltage and current sensor faults. To implement this, we formulate the fault diagnosis problem within the MHE framework, modifying the MHE objective function and constraints to incorporate structural information. To address the computational complexity of MHE optimization, we introduce a hierarchical solution that decomposes the pack-level MHE problem into smaller, more manageable module-level problems, allowing for efficient parallel computation. This approach enables computationally efficient fault diagnosis while requiring significantly fewer sensors. Extensive simulations across various battery pack configurations and fault conditions validate the effectiveness of the proposed method in utilizing structural properties to enhance fault diagnosis.

\balance
\bibliographystyle{IEEEtran}
\bibliography{IEEEabrv,BIB}

\end{document}